\documentclass
[twocolumn,preprintnumbers,superscriptaddress,unsortedaddress]{revtex4}%
\usepackage{amssymb}
\usepackage{amsmath}
\usepackage{graphicx}
\usepackage{dcolumn}
\usepackage{bm}
\usepackage{amsfonts}%
\providecommand{\U}[1]{\protect\rule{.1in}{.1in}}
\providecommand{\U}[1]{\protect\rule{.1in}{.1in}}
\providecommand{\U}[1]{\protect\rule{.1in}{.1in}}
\begin{document}
\title{Noise Induced Intermittency in a Superconducting Microwave Resonator}
\author{Gil Bachar}
\affiliation{Department of Electrical Engineering, Technion, Haifa 32000 Israel}
\author{Eran Segev}
\affiliation{Department of Electrical Engineering, Technion, Haifa 32000 Israel}
\author{Oleg Shtempluck}
\affiliation{Department of Electrical Engineering, Technion, Haifa 32000 Israel}
\author{Steven W. Shaw}
\affiliation{Department of Mechanical Engineering, Michigan State University, East Lansing,
MI 48824-1226 USA}
\author{Eyal Buks}
\affiliation{Department of Electrical Engineering, Technion, Haifa 32000 Israel}
\date{\today }

\begin{abstract}
We experimentally and numerically study a NbN superconducting stripline
resonator integrated with a microbridge. We find that the response of the
system to monochromatic excitation exhibits intermittency, namely,
noise-induced jumping between coexisting steady-state and limit-cycle
responses. A theoretical model that assumes piecewise linear dynamics yields
partial agreement with the experimental findings.

\end{abstract}
\maketitle






Nonlinear response of superconducting RF devices can be exploited for a
variety of applications such as noise squeezing \cite{Movshovich_1419},
bifurcation amplification
\cite{Siddiqi_207002,Castellanos_083509,Tholen_253509} and resonant readout of
qubits \cite{Lee_0609561}. Recently we have reported on an instability found
in NbN superconducting stripline resonators in which a short section of the
stripline was made relatively narrow, forming thus a microbridge
\cite{Segev_160,Segev_57002}. In these experiments a monochromatic pump tone,
having a frequency close to one of the resonance frequencies, is injected into
the resonator and the reflected power off the resonator is measured. We have
discovered that there is a certain zone in the pump frequency - pump amplitude
plane, in which the resonator exhibits limit-cycle (LC) response resulting in
self-sustained modulation of the reflected power. Moreover, to account for the
experimental findings we have proposed a simple piecewise linear model, which
attributes the resonator's nonlinear response to thermal instability occurring
in the microbridge \cite{Segev_096206}. In spite of its simplicity, this model
yields a rich variety of dynamical effects. In particular, as we show below,
it predicts the occurrence of intermittency, namely the coexistence of
different LC and steady state solutions, and noise-induced jumping between
them. In the present paper we study both theoretically and experimentally
noise-induced transitions between different metastable responses. We employ a
1D map to find the possible LC solutions of the system and to find conditions
for the occurrence of intermittency. Experimentally we present measurements
showing both, intermittency between a LC and a steady state, and intermittency
between different LCs. A comparison between the experimental results and
theory yields a partial agreement.

Intermittency is a phenomenon in which a system response remains steady for
periods of time (the laminar phase) which are interrupted by irregular spurts
of relatively large amplitude dynamics (the turbulent phase). It arises in
certain deterministic systems that are near a bifurcation in which a steady
response is destabilized or destroyed \cite{Berge_1984}. This phenomenon also
occurs in noisy systems in which the laminar response has a weak point in its
local basin of attraction and is randomly bumped across the basin threshold,
and then ultimately reinjected back to the laminar state, and the process
repeats. This latter type of bursting behavior, which is relevant to the
present system, is observed to occur in many other systems, including
Rayleigh-Bnard convection~\cite{Ecke_1989}, acoustic
instabilities~\cite{Franck_152}, turbulent boundary layers~\cite{Stone_1989},
semiconducting lasers \cite{Pedaci_036125}, blinking quantum dots
\cite{Kuno_3117}, sensory neurons \cite{Longtin_656} and cardiac tissues
\cite{Chialvo_201}. The presence and level of noise has a significant effect
on all such systems, since perturbations affect the triggering of the system
out of the laminar phase~\cite{Eckmann_1981, Haucke_2090}. The mean duration
times of the laminar phase for a certain class of these systems scales in a
manner that depends on the bifurcation parameter and the noise
level~\cite{Sommerer_1991a, Sommerer_1991b}. A special feature of the present
system is that it exhibits a very sharp transition between two types of
operating states, namely, normal conducting (NC) and superconducting (SC),
which is modeled by equations with discontinuous characteristics. While the
deterministic behavior of such non-smooth systems (at least of low order) is
generally well understood, including local and global
bifurcations~\cite{diBernardo_2007}, the effects of noise in such systems has
not been considered.%
\begin{figure}
[ptb]
\begin{center}
\includegraphics[
height=1.2107in,
width=3.3399in
]%
{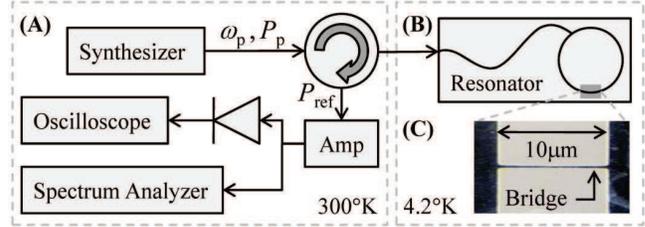}%
\caption{(A) Measurement setup. (B) Schematic layout of the device. (C)
Optical microscope image of the microbridge.}%
\label{Setup}%
\end{center}
\end{figure}

The present experiments are performed using the setup depicted in Fig.
\ref{Setup}$\left(  \mathrm{A}\right)  $ \cite{Segev_1943}. The resonator is
stimulated with a monochromatic pump tone having an angular frequency
$\omega_{\mathrm{p}}$ and power $P_{\mathrm{p}}$. The power reflected off the
resonator is amplified at room temperature and measured by using both, a
spectrum analyzer in the frequency domain and an oscilloscope, tracking the
reflected power envelope, in the time domain. All measurements are carried out
while the device is fully immersed in liquid Helium. A simplified circuit
layout of the device is illustrated in Fig. \ref{Setup}$(\mathrm{B})$. The
resonator is formed as a stripline ring made of Niobium Nitride (NbN)
deposited on a Sapphire wafer \cite{Segev_1943, Chang_1733}, and having a
characteristic impedance of $50%
\operatorname{\Omega }%
$. A feedline, which is weakly coupled to the resonator, is employed for
delivering the input and output signals. A microbridge is monolithically
integrated into the structure of the ring \cite{Saeedkia_510}.

The dynamics of our system can be captured by two coupled equations of motion
\cite{Segev_096206}. Consider a resonator driven by a weakly coupled feedline
carrying an incident coherent tone $b^{\mathrm{in}}=b_{0}^{\mathrm{in}%
}e^{-i\omega_{\mathrm{p}}t}$, where $b_{0}^{\mathrm{in}}$ is constant complex
amplitude and $\omega_{\mathrm{p}}$ is the driving angular frequency.\ The
mode amplitude inside the resonator can be written as $Be^{-i\omega
_{\mathrm{p}}t}$, where $B\left(  t\right)  $ is a complex amplitude, which is
assumed to vary slowly on a time scale of $1/\omega_{\mathrm{p}}$. In this
approximation, the equation of motion of $B$ reads%
\begin{equation}
\frac{\mathrm{d}B}{\mathrm{d}t}=\left[  i\left(  \omega_{\mathrm{p}}%
-\omega_{0}\right)  -\gamma\right]  B-i\sqrt{2\gamma_{1}}b^{\mathrm{in}%
}+c^{\mathrm{in}}, \label{dB/dt}%
\end{equation}
where $\omega_{0}$ is the angular resonance frequency and $\gamma=\gamma
_{1}+\gamma_{2}$, where $\gamma_{1}$ is the coupling coefficient between the
resonator and the feedline and $\gamma_{2}$ is the damping rate of the
mode.\ The term $c^{\mathrm{in}}$ represents an input Gaussian noise, whose
time autocorrelation function is given by $\langle c^{\mathrm{in}%
}(t)c^{\mathrm{in}\ast}(t^{\prime})\rangle=G\omega_{0}\delta\left(
t-t^{\prime}\right)  $, where the constant $G$ can be expressed in terms of
the effective noise temperature $T_{\mathrm{eff}}$ as $G=\left(
2\gamma/\omega_{0}\right)  \left(  k_{B}T_{\mathrm{eff}}/\hbar\omega
_{0}\right)  $. The microbridge heat balance equation reads%
\begin{equation}
C\frac{\mathrm{d}T}{\mathrm{d}t}=2\hslash\omega_{0}\gamma_{2}\alpha\left\vert
B\right\vert ^{2}-H\left(  T-T_{0}\right)  , \label{dT/dt}%
\end{equation}
where $T$ is the temperature of the microbridge, $C$ is the thermal heat
capacity, $\alpha$ is the portion of the heating power applied to the
microbridge relative to the total power dissipated in the resonator
($0\leq\alpha\leq1$), $H$ is the heat transfer coefficient, and $T_{0}=4.2%
\operatorname{K}%
$ is the temperature of the coolant.

Coupling between Eqs. (\ref{dB/dt}) and (\ref{dT/dt}) originates by the
dependence of the parameters of the driven mode $\omega_{0}$, $\gamma_{1}$,
$\gamma_{2}$ and $\alpha$ on the resistance and inductance of the microbridge,
which in turn depend on its temperature. We assume the simplest case, where
this dependence is a step function that occurs at the critical temperature
$T_{\mathrm{c}}\simeq10%
\operatorname{K}%
$ of the superconductor, namely $\omega_{0}$, $\gamma_{1}$, $\gamma_{2}$ and
$\alpha$ take the values $\omega_{0\mathrm{s}}$, $\gamma_{1\mathrm{s}}$,
$\gamma_{2\mathrm{s}}$ and $\alpha_{\mathrm{s}}$ respectively for the SC phase
$\left(  T<T_{\mathrm{c}}\right)  $ of the microbridge and $\omega
_{0\mathrm{n}}$, $\gamma_{1\mathrm{n}}$, $\gamma_{2\mathrm{n}}$ and
$\alpha_{\mathrm{n}}$ respectively for the NC phase $\left(  T>T_{\mathrm{c}%
}\right)  $.

\begin{figure}
[ptb]
\begin{center}
\includegraphics[
trim=0.000000in 0.000000in 0.000000in -0.190366in,
height=1.9458in,
width=2.5365in
]%
{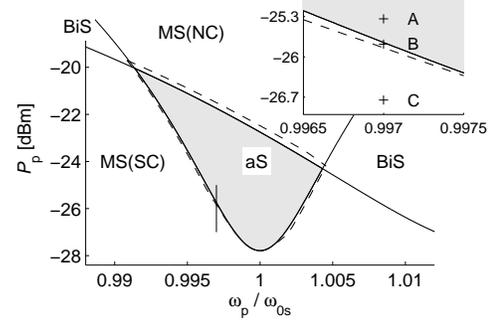}%
\caption{Stability zones in the $\omega_{\mathrm{p}}-P_{\mathrm{p}}$ plane: SC
monostable (MS(SC)), NC monostable (MS(NC)), bistable (biS), and astable (aS)
(gray colored) zones. The region where a stable LC exists is marked with
dashed line. The inset shows the three operating points (A, B and C) at which
the measurements and theoretical analysis shown in Fig. \ref{3casesA} and
\ref{3casesB}\ are done. The following parameters were used in the numerical
simulation: $\omega_{0\mathrm{s}}/2\pi=3.49\operatorname{GHz}$, $\gamma
_{1\mathrm{s}}=0.025\omega_{0\mathrm{s}}$, $\gamma_{2\mathrm{s}}%
=0.060\omega_{0\mathrm{s}}$, $\omega_{0\mathrm{n}}/\omega_{0\mathrm{s}%
}=1.017\omega_{0\mathrm{s}}$, $\gamma_{1\mathrm{n}}=0.25\omega_{0\mathrm{s}}$
and $\gamma_{2\mathrm{n}}=0.60\omega_{0\mathrm{s}}$, $C=15.4\mathrm{fJ}%
/\operatorname{K},$ $H/C=4.62\omega_{0\mathrm{s}}$, $T_{\mathrm{eff}%
}=700\operatorname{K}$. $\omega_{p}=0.997\omega_{0\mathrm{s}}$, $P_{\mathrm{p}%
}=-25.34\operatorname*{dBm}$ (A), $-25.78\operatorname*{dBm}$ (B),
$-26.75\operatorname*{dBm}$ (C). }%
\label{Zones}%
\end{center}
\end{figure}

Solutions of steady state response to a monochromatic excitation are found by
seeking stationary solutions to Eqs. (\ref{dB/dt}) and (\ref{dT/dt}) for the
noiseless case $c^{\mathrm{in}}=0$. The system may have, in general, up to two
locally stable steady states, corresponding to the SC and NC phases of the
microbridge. The stability of each of these phases depend on the corresponding
steady states values $B_{\mathrm{s}}=i\sqrt{2\gamma_{1}}b^{\mathrm{in}%
}/\left[  i\left(  \omega_{\mathrm{p}}-\omega_{0\mathrm{s}}\right)
-\gamma_{\mathrm{s}}\right]  $ and $B_{\mathrm{n}}=i\sqrt{2\gamma_{1}%
}b^{\mathrm{in}}/\left[  i\left(  \omega_{\mathrm{p}}-\omega_{0\mathrm{n}%
}\right)  -\gamma_{\mathrm{n}}\right]  $ [see Eq. (\ref{dB/dt})]. A SC steady
state exists only if $\left\vert B_{\mathrm{s}}\right\vert ^{2}<E_{\mathrm{s}%
}$ where $E_{\mathrm{s}}=H\left(  T_{\mathrm{c}}-T_{0}\right)  /2\hslash
\omega_{0\mathrm{s}}\gamma_{2\mathrm{s}}\alpha_{\mathrm{s}}$, whereas a NC
steady state exists only if $\left\vert B_{\mathrm{n}}\right\vert
^{2}>E_{\mathrm{n}}$ where $E_{\mathrm{n}}=H\left(  T_{\mathrm{c}}%
-T_{0}\right)  /2\hslash\omega_{0\mathrm{n}}\gamma_{2\mathrm{n}}%
\alpha_{\mathrm{n}}$. Consequently, four stability zones can be identified in
the plane of pump power $P_{\mathrm{p}}\propto\left\vert b_{0}^{\mathrm{in}%
}\right\vert ^{2}$ - pump frequency $\omega_{\mathrm{p}}$ (see Fig.
\ref{Zones}) \cite{Segev_096206}. Two are mono-stable (MS) zones (MS(SC) and
MS(NC)), where either the SC or the NC phases is locally stable, respectively.
Another is a bistable zone (BiS), where both phases are locally stable. The
third is an astable zone (aS), where none of the phases are locally stable.%

\begin{figure}
[ptb]
\begin{center}
\includegraphics[
height=3.032in,
width=2.5365in
]%
{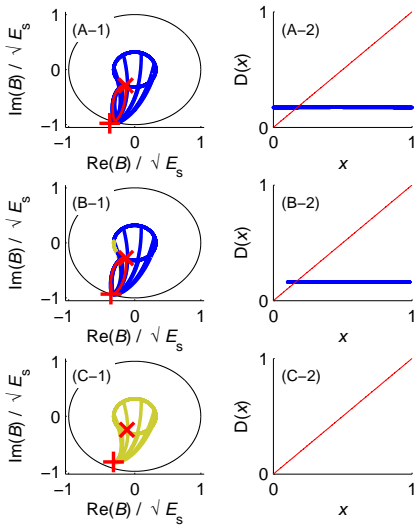}%
\caption{(color online) Resonator's dynamics. Subplots A, B and C correspond
to the three operating points A, B and C respectively which are marked in the
inset of Fig. \ref{Zones}. In subplot (A) only a LC is locally stable, in
subplot (B) intermittency between a LC and a SC steady state occurs, whereas
only a SC steady state is locally stable in subplot (C). In panels (A-1),
(B-1) and (C-1), which show the time evolution in $B$ plane, a plus sign
labels $B_{\mathrm{s}}$ and a cross sign labels $B_{\mathrm{n}}$. These points
are shown for reference and correspond to fixed points of the dynamics only
when they exist in their respective domains, as defined in the text.
Trajectories that return to the inner circle $\left\vert B\right\vert
^{2}=E_{\mathrm{n}}$ are colored in blue (dark), and trajectories that end at
$B_{\mathrm{s}}$ are colored in yellow (gray). Panels (A-2), (B-2) and (C-2)
show the corresponding 1D maps.}%
\label{3casesA}%
\end{center}
\end{figure}

The task of finding LC solutions of Eqs. (\ref{dB/dt}) and (\ref{dT/dt}) can
be greatly simplified by exploiting the fact that typically $\gamma\ll H/C$ in
our devices, namely, the dynamics of the mode amplitude $B$ [Eq.
(\ref{dB/dt})] can be considered as slow in comparison with the one of the
temperature $T$ [Eq. (\ref{dT/dt})]. In this limit one finds by employing an
adiabatic approximation \cite{Segev_096206} that the temperature $T$ remains
close to the instantaneous value given by $T_{\mathrm{i}}=T_{0}+2\hslash
\omega_{0}\gamma_{2}\left\vert B\right\vert ^{2}/H$ for most of the time
except of relatively short time intervals (on the order of $C/H)$ right after
each switching event between the SC and NC phases. Consequently, as can be
seen from the example trajectories shown in Fig. \ref{3casesA} (A-1),
transitions from SC to NC phase occur near the circle $\left\vert B\right\vert
^{2}=E_{\mathrm{s}}$, whereas transitions from NC to SC phase occur near the
circle $\left\vert B\right\vert ^{2}=E_{\mathrm{n}}$.%
\begin{figure}
[ptbptb]
\begin{center}
\includegraphics[
trim=0.000000in 0.000000in 0.000780in 0.000000in,
height=2.0695in,
width=2.5374in
]%
{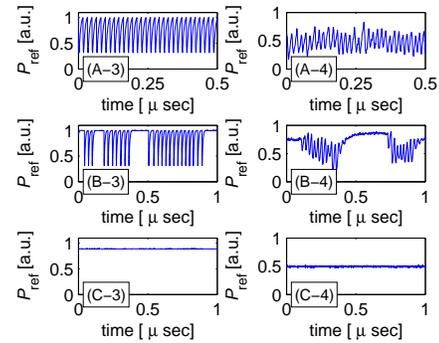}%
\caption{Numerical [panels (A-3), (B-3) and (C-3)] vs. experimental [panels
(A-4), (B-4) and (C-4)] time traces for the three operating points A, B and C
respectively.}%
\label{3casesB}%
\end{center}
\end{figure}

The important features of the system's dynamics can be captured by
constructing a 1D map \cite{Strogatz_Nonlinear}. Consider the case where
$E_{\mathrm{n}}<E_{\mathrm{s}}$ and the amplitude $B$ lies initially on the
circle $\left\vert B\right\vert ^{2}=E_{\mathrm{n}}$, namely $B=\sqrt
{E_{\mathrm{n}}}e^{2\pi ix}$ where $x\in\lbrack0,1]$. Furthermore, assume that
initially the system is in the SC phase, namely, $T<T_{\mathrm{c}}$ and
consequently $B$ is attracted towards the point $B_{\mathrm{s}}$. The 1D map
$D\left(  x\right)  $ is obtained by tracking the time evolution of the system
for the noiseless case ($c^{\mathrm{in}}=0$) until the next time it returns to
the circle $\left\vert B\right\vert ^{2}=E_{\mathrm{n}}$ to a point
$B=\sqrt{E_{\mathrm{n}}}e^{2\pi iD\left(  x\right)  }$ where $D\left(
x\right)  \in\lbrack0,1]$. In the adiabatic limit this can be done using Eq.
(\ref{dB/dt}) only [without explicitly referring to Eq. (\ref{dT/dt})] since
switching to the NC phase in this case occurs when the trajectory intersects
with the circle $\left\vert B\right\vert ^{2}=E_{\mathrm{s}}$. Note that in
the aS zone of operation all points on the circle $\left\vert B\right\vert
^{2}=E_{\mathrm{n}}$ return back to it after a finite time. However, this is
not necessarily the case in the other stability zones. Therefore, we restrict
the definition of the 1D map $D\left(  x\right)  $ only for points on the
circle $\left\vert B\right\vert ^{2}=E_{\mathrm{n}}$ that eventually return to
it. Other points will have a trajectory that ends at a steady state (NC or SC).

Any fixed point of the 1D map, namely a point for which $D\left(
x_{0}\right)  =x_{0}$, represents a LC of the system. The LC is locally stable
provided that $\left\vert \mathrm{d}D/\mathrm{d}x\right\vert _{x=x_{0}}<1$
\cite{Strogatz_Nonlinear}. The region in the $\omega_{\mathrm{p}}$ -
$P_{\mathrm{p}}$ plane in which a locally stable LC solution exists was
determined using the parameters of our device and it is marked with dashed
line in Fig. \ref{Zones}.

Figure \ref{3casesA} shows noiseless behvior of the resonator for the three
operating points A, B and C, which lie near the border between the aS region
and the MS(SC) one, and are marked in the inset of Fig. \ref{Zones}. Figure
\ref{3casesB} shows a comparison of experimental data and numerical simulation
for these operating points. The sample parameters used in the numerical
simulations and are listed in the caption of Fig. \ref{Zones}, were determined
using the same methods detailed in Ref. \cite{Segev_096206}.

Subplot (A) shows the behavior at operating point A, which lies inside the aS
zone. In panel (A-1) sample trajectories in $B$ plane are shown. The resultant
1D map, which is plotted in panel (A-2), has a single fixed point
corresponding to a single locally stable LC. The time evolution seen in panel
(A-3) was obtained by numerically integrating the coupled stochastic equations
of motion (\ref{dB/dt}) and (\ref{dT/dt}). The trace is then compared to
experimental data taken from the same working point [panel (A-4)].

At operating point B [see Fig \ref{3casesA} and \ref{3casesB} subplot (B)]
coexistence of a LC and a SC steady state occurs. The LC corresponds to the
locally stable fixed point of the 1D map seen in panel (B-2). On the other
hand, all initial points on the circle $\left\vert B\right\vert ^{2}%
=E_{\mathrm{n}}$ that never return to it evolve towards the SC steady state
$B_{\mathrm{s}}$. Numerical time evolution and experimental data shows
noise-induced transitions between the two metastable solutions [panels (B-3)
and (B-4) respectively]. At operating point C [see Fig \ref{3casesA} and
\ref{3casesB} subplot (C)] the LC has been annihilated by a
discontinuity-induced bifurcation~\cite{diBernardo_2007} and consequently only
steady state response is observed.%
\begin{figure}
[ptb]
\begin{center}
\includegraphics[
trim=-0.016063in 0.000000in 0.016063in 0.000000in,
height=1.177in,
width=2.2563in
]%
{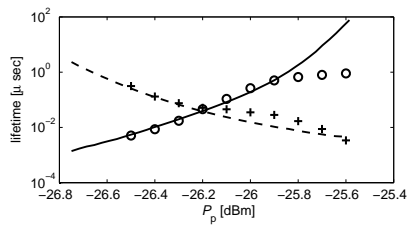}%
\caption{Experimental data of life time of LC (circles) and steady state
(pluses) compared to numerical prediction (solid and dashed lines
respectively). Mesurements above $-25.8\operatorname*{dBm}$ saturate to
$1\mu\sec$ as this is the maximal mesuerment time.}%
\label{LifeTime}%
\end{center}
\end{figure}

To further study noise-induced transitions we fixed $\omega_{\mathrm{p}}$ and
vary $P_{\mathrm{p}}$ starting from MS(SC) zone $P_{\mathrm{p}}%
=-26.7\operatorname*{dBm}$ to the aS zone $P_{\mathrm{p}}%
=-25.6\operatorname*{dBm}$ [vertical line in Fig. (\ref{Zones})], and took
$1\mu\sec$ long time traces of $P_{\mathrm{ref}}$ [similar to those seen in
Fig. \ref{3casesB} (A-4), (B-4) and (C-4)]. The average lifetime of both LC
and SC steady state, namely, the average time the system is in one solution
before making a transition to the other one, were determined from these
traces. This data, compared to numerical simulation prediction is shown in
Fig. \ref{LifeTime}.

\begin{figure}
[ptb]
\begin{center}
\includegraphics[
height=1.9242in,
width=2.5374in
]%
{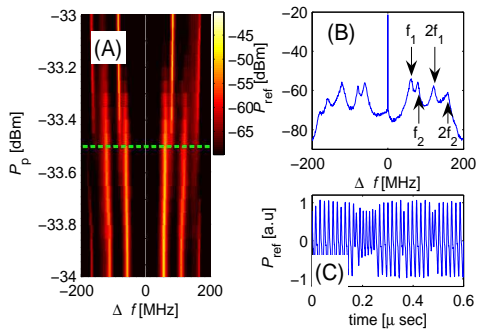}%
\caption{Experimental demonstration of intermittency between two LCs. Panel
(A) shows the reflected power $P_{\mathrm{ref}}$ as a function of the offset
frequency $\Delta f=(\omega-\omega_{p})/2\pi$ and the pump power
$P_{\mathrm{p}}$. Panel (B) shows a cross section of panel (A) which is
indicated by a dashed line. The frequencies $f_{1}$ and $f_{2}$ of the two LCs
are marked. Panel (C) shows a time trace of the reflected power taken at the
same value of $P_{\mathrm{p}}$. In this experiment $\omega_{p}/2\pi
=6.61\operatorname{GHz}$}%
\label{2LC}%
\end{center}
\end{figure}

In another experiment using a similar device we observe intermittency of two
different LCs (see Fig. \ref{2LC}). Panel (A) shows spectrum analyzer
measurement of the reflected power as a function of the pump power
$P_{\mathrm{p}}$. Two distinct LCs having frequencies $f_{1}\simeq
60\operatorname{MHz}$ and $f_{2}\simeq80\operatorname{MHz}$ are observed. For
$P_{\mathrm{p}}<-33.5\operatorname*{dBm}$\ only a LC at frequency $f_{1}$ is
visible. In the range $-33.55\operatorname*{dBm}<P_{\mathrm{p}}%
<-33.35\operatorname*{dBm}$ both LCs\ are seen, whereas for high pump power
$P_{\mathrm{p}}>-33.5\operatorname*{dBm}$ only a LC at frequency$\ f_{2}$ is
seen. Panels (B) and (C), demonstrates the behavior in the intermediate
region, where both LCs are observed in frequency and time domain respectively
for $P_{\mathrm{p}}=-33.5\operatorname*{dBm}$ [marked by a dashed line in
panel (A)].

In general, intermittency of two (or more) different LCs can be theoretically
reproduced using our simple model. However, we were unable to numerically
obtain this behavior without significantly varying some of the system's
experimental parameters. This discrepancy between experimental and theoretical
results suggest that a further theoretical study is needed in order to develop
a more realistic description of the system.%

We thank Mike Cross, Mark Dykman, Oded Gottlieb and Ron Lifshitz for valuable
discussions. This work was supported by the ISF, Deborah Foundation, Poznanski
Foundation, RBNI and MAFAT.

\bibliographystyle{apsrev}
\bibliography{Eyal_Bib}

\end{document}